C/Asteroids/ActiveAsteroids/PSS/25NewLowMOD/160914/170119

# "Secular light curves of 25 members of the Themis Family of asteroids, suspected of low level cometary activity"

Ignacio Ferrín, Martín Pérez, Juan Rendón
FACom and Sun, Earth and Planetary Physics Group
Institute of Physics
Faculty of Exact and Natural Sciences
University of Antioquia
Medellín, Colombia 00051000
ignacio.ferrin@udea.edu.co

**Abstract**

From 1996 to 2015 sixteen main belt asteroids were discovered exhibiting cometary activity (less than one per year), all of them during searches at the telescope. In this work we will explore another way to discover them. We reduced 192016 magnitude observations of 165 asteroids of the Themis family, using data from the astrometric-photometric database of the Minor Planet Center, MPCOBS, and measuring the absolute magnitudes from the phase plots. 25 objects of 165 (15.2%), exhibited bumps or enhancements in brightness that might indicate low level cometary activity. Since activity repeats at the same place in different orbits and in many occasions is centered at perihelion, activity might be due to water ice sublimation. As of September 2016, there are 717768 asteroids listed in the Minor Planet files. If we assume that we do not have any false positives and the above percentage can be extrapolated to the whole Main Belt, the number of potentially active asteroid gets to the very large number of ~111.000. This number is much larger than the ones predicted in previous surveys and indicates one of three scenarios: A) there are many false positives in our detections and the real number of active asteroid is much smaller than we found. This would be tantamount to saying that the MPC astrometric-photometric database is only astrometric and not photometric, invalidating absolute magnitude determinations based on this database. B) The location of active asteroids is restricted to the Themis family and an extrapolation to the whole belt is not possible. Or C) there are few false positives in our candidates and the main belt actually contains many low level active asteroids undetected by current surveys. Case C) would imply that the main belt is not a field of bare rocks but a graveyard of extinct comets, changing our current paradigm of the main belt. So it is of the outmost importance to verify *observationally* our candidates, and determine which of these scenarios is valid.

**1. Introduction and Method of Detection**

Up to 2015 sixteen main belt asteroids have been discovered exhibiting cometary activity. The first one was 133P/Elst-Pizarro observed in 1996 (CIAU 6456), the last two ones 62412 (Sheppard and Trujillo, 2015) and P/2015 X6 PANSTARRS (Lilly and Weryk, 2015). These main belt comets (MBCs) are important because they would represent a new reservoir of comets, if they were created in situ, and they might change our understanding of the origin, evolution and structure of the main belt of asteroids (Jewitt et al., 2015).

These objects were discovered at the telescope, searching for asteroids with coma or tails. This method of discovery is very inefficient, since less than one object has been discovered per year, and consumes large amounts of telescope time (Hsieh et al., 2015; Gilbert and Wiegert, 2009; Sonnett et al., 2011; Waszczak et al., 2013). In this work we explore another way to discover them.

We will use the methodology of the Secular Light Curves of comets, SLCs (Ferrín, 2010), to calculate the absolute magnitude of the object from the equation

$$m(1,1,0) = m(\Delta, R, \alpha) - 5 \log \Delta \cdot R - \beta \cdot \alpha \qquad (1)$$

using the Minor Planet Center database (http://www.minorplanetcenter.net/db_search), and to plot it vs time to perihelion, $t - T_q$. $m(\Delta,R,\alpha)$ is the observed magnitude, $\Delta$ the asteroid-Earth distance, $R$ the asteroid-Sun distance, $\alpha$ is the phase angle, $\beta$ the phase coefficient and $T_q$ the time of perihelion. The linear solution for the phase coefficient is justified since the phase angle never exceeds 23°. After reducing 105 asteroids of the Themis family, we found a mean value $<\beta> = 0.050 \pm 0.02$. The absolute magnitude of an inactive asteroid must be a constant vs time or place in the orbit, while an active asteroid will exhibit magnitude bumps or enhancements. We examined only those asteroids that have rotational periods determined, listed in the LCDB database (Warner et al., 2009), in order to study if the rotational period affects cometary activity (Jewitt et al., 2014). We find negative evidence of such a correlation, evidence presented later on in Table 3 and in Figures 16-17.

The decision to investigate the Themis family was based on the fact that this family contains the largest number of known active comets and water ice surface contaminated asteroids than any other main belt family, as deduced from Table 1. Infrared spectroscopic measurements have shown that some members contain water ice signatures on their surfaces (Campins et al., 2010; Hargrove et al., 2015; Rivkin and Emery, 2010; Fornasier et al., 2016).

The Themis family of asteroids is a well-known, statistically stablished group, discovered by Hirayama (1918). It has been identified as a family by many authors. According to Zappala et al. (1995), it contains more than 550 members, while Nesvorny (2012), identified more than 5000.

The complete evidence suggest that the Themis family is a significant reservoir of water in the outer main belt (Table 1).

Water ice seems to be present in other asteroids like (65) Cybele, so it is not restricted to the Themis family (Licandro et al., 2011). But it has in common with the Themis family being a member of the outer main belt.

## 2. Comparison light curve: 107P Wilson Harrington

The candidates to active asteroids presented in this work exhibit low activity that resembles that of comet 107P/Wilson-Harrington, a well-studied and characterized object (Ferrín, 2012). The reduction procedure used in this work is the same used to reduce this comet.

The object was discovered in 1979 as asteroid 1979VA. When an orbit was determined it received the number 4015. Soon after it was seen as an active comet in two old Palomar Observatory survey plates, the red and the blue thus receiving the denomination 107P/Wilson-Harrington. There is no other observation in the literature registering the object as active. During the whole 2009 apparition, the comet was stellar in appearance, and the activity raised above the nuclear magnitude by only -0.70±0.10 magnitudes, while the duration was ~69±5 days (Figure 1). Since there was a magnitude enhancement, the activity must have been contained inside the seeing disc, and no coma or tail were detected.

This result shows that enhancements of less than one magnitude and durations less than a hundred days are easily detected. Next we applied the SLC methodology to 165 members of the Themis family, finding that 25 objects exhibit unexplained bumps and enhancements that might be interpreted as low level cometary activity. In many instances the activity is centered at perihelion or past perihelion, confirming this suspicion.

## 3. Object Detection and data reduction

A plot of $m(1,1,0)$ vs $t - T_q$ shows significant vertical dispersion that we interpreted as due to the fact that MPCOBS is an astrometry database, and small photometric apertures are used that do not extract the whole flux from the asteroid. However, a *well-defined envelope* appears in all negative plots. This is being interpreted as due to apertures that extract the whole flux from the asteroid. In stellar photometry it is well known (Howell, 1992; Da Costa, 1992), that the maximum signal to noise ratio is achieved with measuring apertures about ~0.75xFWHM, typically used in astrometry, while to extract the complete flux from the object at least ~5xFWHM are needed. The difference represents a data point displace downward by ~3 magnitudes, which is observed in many of our plots. *Thus we take the envelope as the correct interpretation of the SLC.*

When we take the envelope of the dataset, we take the envelope of the rotational light curve too. The rotational light curve can be modeled as a sine wave. If a sine wave is sampled at equal intervals, the distributions shows maxima at the extremes (Figure 2), maximum and minimum brightness. The maximum brightness helps to make the envelope sharp. To make the envelope even sharper, we reduced all magnitude observations to the V-band using the transformation equations of Jordi et al. (2008). In this way the uncertainty of the envelope was reduced to ~±<0.20 magnitudes, our detection limit (Figure 3). This value is confirmed by the envelopes of the light curves of the inactive asteroids (SI).

There are a few data points that lie above the envelope and that encompass only a few days. We interpreted these as due to contamination by nearby stars in congested fields, cosmic rays, CCD defects or catalog errors. That this may be the case, can be confirmed by the fact that there are also perturbations on the astrometry by background objects (Ivantsov et al., 2015).

For asteroids with no activity (a bare nucleus), the absolute magnitude is flat and independent on time or location on the orbit by definition (confirmation, see Inactive Asteroids, in the Supplementary Information, SI). Objects with activity, on the other hand, show localized bumps above the envelope, many of them preferentially near perihelion.

The examination of the SLC, m(1,1,0) vs t-Tq, allows us to declare an object positive (+), or negative (-) if: a) the activity appears at several apparitions at the same place in the orbit. b) The activity appears during a time span of 100 d or more (our detection limit). c) The amplitude of the SLC, $A_{SEC}$ = m(1,1,0) – m(MAX), is larger than ~0.2 magnitudes (our detection limit, after filters correction, Figure 3).

Some plots separate de data by apparition year to search for recurrent activity. Recurrent activity at different returns is currently considered the most reliable indicator of sublimation driven activity (Hsieh et al., 2015).

The absolute magnitude can be determined using two methods: a) the phase plot extrapolated to zero phase (Figure 4), and b) the histogram of m(1,1,0) values (Figure 5). For 105 objects we found m(1,1,0)(phase) – m(1,1,0)(histogram) = ±0.09±0.03 magnitudes, thus the agreement is excellent. The phase plots are all very similar and uneventful and thus are not shown in the SI.

Some phase plot showed the existence of the opposition effect. The observations that showed this effect were removed. Thus the bumps and enhancements are not due to the opposition effect.

We are interested in determining the absolute magnitude m(1,1,0), and we equal it to the middle of the rotational light curve, whose maximum amplitude Arot, is extracted from the LCDB catalog (Warner et al., 2009). However, when we take the envelope of the dataset, we take the envelope of the rotational light curve too, Env, because the rotational periods are of the order of hours and thousands of cycles are observer per apparition. They are related thus:

m(1,1,0) = Env + 0.5 Arot

Δm = m110(MAX) – Env = m110(MAX) + 0.5 Arot - m(1,1,0)

$A_{SEC}$ (True) = m110(MAX) - m(1,1,0) = Δm -0.5 Arot     (2)

Where Δm is the amplitude of the enhancement and it is negative, m110(MAX) is the peak magnitude, and $A_{SEC}$ is the true amplitude of the enhancement.

In an investigation of this nature, there will be false positives and false negatives in cometary activity. So there is the possibility that a fraction of our candidate objects may not be active or that a fraction of our negative objects may be active. It is difficult to detect activity below ~<100 days of duration. However all objects were reduced and examined thrice to reduce this possibility, so we are confident in our detections. We warn that we do not discard the possibility of having statistical flukes or that large rotational amplitudes may influence the results.

Following these rules, we processed 192016 photometric magnitude observations of 165 asteroids members of the Themis Family, using the methodology of the SLCs of comets, and searching for cometary activity. All of them were taken from the LCDB database (Warner et al., 2009), finding 25 (15.2%) candidates with probable positive cometary activity (+), 106 (64.2%) candidates with negative activity (-), 31(18.8%) candidates with evidence of large obliquity, and 3 (1.8%) candidates with evidence of being a nearby, contact binary, or very elongated object.

Figure 6 shows 4 non-active asteroids all of them belonging to the Themis family, to get acquainted with the SLC of an asteroid with no activity. We see that the envelope is rather sharp within ~±<0.20 magnitudes, our 3-$\sigma$ limit of detection. There are no bumps or large deviation for these envelopes. Thus these objects have been classified as inactive or negative (-). On the other hand, Figure 7 show 4 possibly active asteroids. All of them show localized bumps or deviations from the envelope, and they have been classified as candidates to active or positive (+). The SLCs of all 165 asteroids are presented in the Supplementary Information, SI.

Figures for positive asteroids list 5 parameters, the onset of activity, Ton; the lag time of maximum light with respect to perihelion, LAG; the offset of activity, Toff; the active interval, $\Delta t$ = Toff – Ton, all of them in units of days; the amplitude of the activity in magnitudes, $\Delta m$, negative because it is a brightening (Equation 2). These parameters are useful at the time of observing these objects at the telescope, and to quantify the activity. From these Figures several histograms can be elaborated.

## 4. Statistical Significance

It is important to establish the statistical significance of the detections. The SLCs of the 165 asteroids presented in the SI, show that the vertical dispersion of the envelope of the inactive objects, is ~±<0.20 magnitudes. Alternatively, the statistical significance can be established from the amplitude of the ratio = Asec / (0.5 rotational amplitude). In 88% of our cases this ratio is larger than 3-$\sigma$. However this dispersion describes only the vertical axis.

The statistical significance of the horizontal axis, t-Tq, can be calculated from the horizontal distribution of the data points. A $\chi^2$–analysis can be used to calculate the p-value, a measure of the randomness of the horizontal distribution. If we do this calculation for asteroid 1342 Yvette we find p=0.0014 or 0.14%, which means that the distribution with the enhancement can hardly be random. The tool to do this calculation can be found here http://www.quantpsy.org/chisq/chisq.htm . Very small percentages are found for all 25 objects detected. However, as stated earlier in this manuscript, in a research of this nature there is always the possibility of having false positives, false negatives and statistical flukes, and we are well aware of this possibility.

**5. Influence of the rotational light curve on the secular light curve**

We define the shape of the asteroid as an ellipsoid of revolution of semi-axis a:b:c such that a>b>c. Then the rotational light curve of the asteroid observed at different longitudes in the orbit (different times), is determined by the parameters a:b:c, the aspect angle A (the angle between the polar axis and the Earth), the obliquity (the angle of the pole and the ecliptic), and the phase angle α (the angle Earth-asteroid-Sun). Since asteroids in the main belt observed from Earth never exceed ~23° of phase angle (confirmation Figure 4), the influence of this parameter is negligible. Figure 4 of Dunlap (1972) shows that the influence of the obliquity is small if its value is > 50° (90° obliquity means that we are observing the asteroid in the equatorial plane). The same happens for the aspect angle (90° aspect angle means that we are observing the asteroid in the equatorial plane). The projected surface area of the ellipsoid is given by the following two equations (Pospieszalska, Surdej and Surdej, 1985):

$$\text{Smax} = \pi abc\ [\ \sin^2(A)/b^2 + \cos^2(A)/c^2\ ] \qquad (3)$$

$$\text{Smin} = \pi abc\ [\ \sin^2(A)/a^2 + \cos^2(A)/c^2\ ] \qquad (4)$$

and the observed magnitude is given by mag = constant – 2.5 Log S.

For A → 0° (pole on orientation) both equations tend to S = πab . For A → 90° (equator on orientation) Smax → πac and Smin → πbc . If the ellipsoid has b = c, then the A → 0° and the A → 90° orientations have the same maximum value (Figure 8).

If the obliquity (and thus the aspect angle) is large, we are always observing the asteroid equator on, and the rotational light curve is approximately a sinusoid that collapses to amplitude zero at the pole-on configuration.

The largest influence of the rotational light curve on the secular light curve is when the object has a large a/b value and the obliquity (and thus the aspect angle) are large.

The best way to show the influence of the rotational light curve on the secular light curve is by showing an example. In Figure 9 we show the case of 7454 Kevinrighter. The signature we are looking for has to have two maxima and two minima in the secular light curve, from (–) aphelion to perihelion and again to (+) aphelion, separated by about half the orbital period. If these two maxima and two minima signature do not appear, that is, if the secular light curve looks flat, we conclude that the obliquity of the object is > 60° and does not influence the secular light curve significantly (Dunlap, 1972).

We have examined the 165 light curves reduced in this work and we have found 31 objects (18.8%) were the rotational light curve influences the secular light curve significantly (Table 2). Of course these objects have to be confirmed by more precise photometry and deserve further study.

We also found evidence that 1576 Faviola, 1633 Chimay and 1778 Alfven might be nearby, contact binaries or very elongated objects. Figure 10 shows the light curves of these three asteroids.

In conclusion, the plots of m(1,1,0) vs t – Tq serve to deduce the parameters of the object (a/b, a/c, the obliquity and the longitude of the pole) if the obliquity is large. However, since this is beyond the scope of this work, we leave this analysis for another occasion.

We would like to study the properties of our positive candidates, and to compare them with the main belt comets discovered up to now. Thus to proceed, we will assume that we do not have any false positives.

## 6. Results: Systematic difference in the absolute magnitude

Figure 11 shows the absolute magnitudes of our candidate objects derived from the generated plots vs the absolute magnitude given by MPCOBS. We see that there is a systematic difference amounting to +0.32±0.19 magnitudes with our absolute magnitudes being brighter, and that some values differ by almost 1 magnitude. This affects the calculated diameter of the objects. Juric et al. (2002), found a similar result. They reported a systematic error of about ~0.4 magnitudes in the MPC's absolute magnitudes in good agreement with our value. Pravec et al. (2012) found the same result, with differences of up to +0.4 to +0.5 magnitudes for the same database.

## 7. Results: Amplitude of the SLC

Figure 12 shows the number of objects vs the amplitude of the activity, $A_{SEC}$ from Equation 3, negative because it is an enhancement. The main belt comets discovered up to now (1996-2014) (black rectangles), have a large amplitude of the SLC, thus they are easy to detect. The 25 objects discovered in this work are faint and difficult to find observationally because they do not exhibit a tail or coma. The SLCs of these low activity comets resemble that of the well-studied comet 107P/Wilson-Harrington (Figure 1) and appear in the SI.

## 8. Results: Maximum of activity vs time to perihelion

Figure 13 shows the number of objects vs LAG, the time of maximum activity with respect to perihelion. We find that the active asteroids are active all along the orbit, but there is a concentration of objects near and after perihelion. This is consistent with activity due to sublimation from an ice layer below the surface. The LAG time may be interpreted as the time it takes the thermal wave to penetrate to the layer. This result implies that active asteroids, MBCs, are not ice coated rocks, but that the ice lies deep inside the object below a layer of low thermal conductivity dust.

## 9. Results: Duration of activity

Figure 14 shows the number of objects vs duration of activity. We see that activity extends for more than a year for many objects. Our lower limit of detection is ~100 d and we cannot detect reliably anything below that value. Many asteroids show continuous activity during one year or more. We may have some false negatives there.

**10. Results: Centrifugal force activation mechanism**

We selected the LCDB database because it contains rotational periods for asteroids and this will allow us to test the hypothesis that the activity of the asteroids is due to rotational instability (Jewitt et al., 2014). An object will be near the breakup limit if it has a rotational period near the critical rotation period, $P_{CRITICAL}$, at which the centripetal force equals the gravitational force. It is given by Jewitt et al. (2014) as:

$$P_{CRITICAL} = (1000 \; 3.3 h^2 \; a/b \; \rho)^{1/2} \qquad (5)$$

where $\rho$ the density of the object. For $1 < a/b < 1.5$ and density $1000 < \rho < 2000$ kg/m$^3$ we find $2.33 < P_{CRITICAL} < 4.04$ hours.

The objects with fast rotational periods are collected in Table 3 and plotted in Figure 15. This Figure shows that there is an excess of objects in the 3-4 hours bin. In Figure 16 we test the correlation duration of activity vs centrifugal/gravitational force = $C/\rho P^2$, where C is a constant, $\rho$ is the density and P the rotational period. We once again see three objects detached from the main concentration. However their duration of activity is the same as for other Themis members. There is no evidence of correlation duration vs centrifugal/gravitational force. Several objects with fast rotation do not exhibit activity (bottom of the plot). In Figure 17 we test the correlation amplitude of activity vs centrifugal/gravitational force. The three Themis members remain detached from the main concentration, however their amplitudes are the same as the other members. Once again we do not find evidence of correlation between these two parameters. Additionally, the fact that other Themis members exhibit fast rotations but no activity suggests that fast rotation is not a sufficient condition for activity.

We conclude that there is no evidence of a correlation of fast rotation with activity. However the question remains, why is there an excess of fast rotators? Masiero et al. (2009) also found an excess. We want to advance the hypothesis that *fast rotators might the remnant of old comets that have suffered spin up due to jets, but that did not split because there is a monolith at the center.*

**11. Location of objects in the $A_{SEC}$ vs D diagram**

In the phase space $A_{SEC}$ vs Diameter (Figure 18), the active members of the Themis family plot in the lower region of the diagram with $A_{SEC}$ < 1 magnitude (Figure 12). This is the *graveyard*. Since some members are larger than 100 km in diameter, they also plot in the *goliath region*. $A_{SEC}$ measures the activity of the asteroid with respect to the nucleus absolute magnitude. $A_{SEC}$ = 0 corresponds to a bare nucleus with no activity.

**12. Number of active objects in the Main Belt**

(a) Gilbert and Wiegert (2009) were the first to attempt to find MBCs at the telescope searching 12390 asteroids. They looked for comas and tails. None where found active.

Later on they updated their search (Gilbert and Wiegert, 2010), searching an additional 13802 objects and again none were found active. From this result they determined an upper limit of active objects of 40±18 in the entire asteroidal belt.

(b) Sonnett et al. (2011) also placed limits on the number of active asteroids. No active MBC were found, but they report an upper limit of 2000 MBC per $10^6$ objects. Since the current population of objects in the Main Belt is 714.825 as of June, 2016, (Marsden et al., 1994), it is possible to calculate that the upper limit on the number of active MBCs should be ~1400.

However, they found an interesting result: about 5% of the main belt objects may exhibit low-level activity that cannot be detected for individual objects, but shows up in the statistical result. This hypothesis is confirmed in our investigation, but the percentage is larger (15.2%).

(c) More recently Waszczak et al. (2013) searched the sky using the Palomar Transient Factory data. From 220.000 photometric observations they identified 1577 comet candidates. From these observations they calculate 33 active MBC per $10^6$ main MB asteroids. Using the same population cited above, it is possible to calculate that the number of active MBC predicted by these authors would be ~23.

(d) Sheppard and Trujillo (2015) placed an order of magnitude constrain in the number of possible active asteroids in the main belt. They detected 15.000 MBA and one active asteroid, giving a rate of 1/15.000. Using the same population cited above, it can be calculated that according to their results there would be ~47 active MBCs.

(e) After setting magnitude limits of $12 < Hv < 19.5$ for the absolute magnitude, Hsieh et al. (2009) calculate that there is a total expected population of ~140 MBCs and an upper limit of ~240 MBCs.

(g) In this work we found 25 candidates to active objects in 165 members of the Themis family of asteroids, or 15.2%. Scaling to the whole belt would give a very large number. This may be due to the existence of false positives, but there is also the possibility that a fraction of the candidates might be real, suggesting the existence of many more active objects than previous estimates. So the question is, why the large difference?

**13. Failure of previous detection methods**
Detection of comas and tails on the sky images is prone to failure due to the observational result, that activity up to 3 magnitudes above the nucleus magnitude produces a coma so small that it is completely contained within the seeing disk (Ferrín, 2012). Consider that comet 28P/Neujmin 1 exhibited a coma with V(observed)-V(nuclear)= -3.2 magnitudes, where V(nuclear) is the nucleus magnitude. Comet 133P/Elst-Pizarro did not exhibit a coma with V(observed)-V(nuclear)= -2.8 magnitudes. Comet 2P/Encke did not exhibit a coma at aphelion with V(observed)-V(nuclear)= -2.4 magnitudes. And comet 107P/Wilson-Harrington did not exhibit a coma with V(observed)-V(nuclear)= -0.70 magnitudes during its 2009 return. Thus there seems to be an intermediate value at which the coma disappears hidden inside the seeing disk.  This quantity is defined as the Threshold Coma Magnitude and its estimated value based on the above examples is TCM ~ -3.0±0.3 magnitudes, and of course, it is dependent on the seeing value, dependence that has not yet been quantified.  If the total minus nuclear magnitude is less than the TCM, the comet may be active, but the coma is contained inside the seeing disk and cannot be detected by imaging.  Consider that at 1.5 AU (the minimum distance of a main belt asteroid) a typical seeing disk of 2.0" corresponds to 2184 km, while the effective radius of many asteroids does not exceed 50 km, thus there is plenty of space for an undetected coma.

This explains why it is so difficult to detect active asteroids at the telescope and why estimates of their existence give so small values.   On the other hand, exploring the SLCs is a fast, reliable and sensitive method.  Sixteen active asteroids were discovered in a time span of 18 years of observations by the former method, while we discovered 25 candidates to active asteroids in one month of learning and data reduction.  Certainly, there are many more active asteroids to be discovered in the belt, thus this research line should be continued.  The existence of many objects with low level cometary activity is expected on physical grounds.

**14.  Themis Family and the snow line**

The age of a family could play an important role in cometary activity.  Younger asteroid families could have more water-rich materials than the older ones, because of the shorter exposure time to the solar radiation.  However this hypothesis does not seem to be sustained in this case.  The age of the Themis family has been estimated at ~2.5±1.0 Gyr by Broz et al., (2013).  It was formed from a large-scale catastrophic collision with a ~270 km to ~400 km parent body (Broz et al., 2013; Marzari et al., 1995; Durda et al., 2007).  Thus it is surprising to find so much water in the Themis family (confirmation Table 1).  We suspect that activity may not depend so much on age as on distance to the Sun.  The Themis family is in the outer region of the asteroidal belt with mean semi-major axis ~3.14±0.09.   (65) Cybele, an asteroid with indication of water ice on its surface (Licandro et al., 2011), is also an outer main belt asteroid (a=3.43AU).

Most probably, the large water content of Themis family asteroids may be related to the location of the snow line.  Hayashi (1981) locates the snow line at 2.7 AU with a condensation temperature of 170° K.  Podolak and Zucker  (2010), place it at 3.2 AU and 143° K, while Martin and Livio (2012) place it at 3.1 AU, with a temperature of 150° K for micron-size grains, and 200° K for km-size bodies (D'Angelo and Podolak, 2015).  Lecar et al. (2006) calculate the location of the snow line in the solar system between 1.6 and 2.7 AU and temperatures of 149° to 185° depending on the model assumptions.  They also found that moving the snow line past 3 AU requires unrealistic values of some parameters.  These results reinforce the idea that the Themis family of asteroids is beyond the current solar system estimates of the snow line, explaining the richness of water ice components in this family.

**15. Conclusions**

**(a)** We reduced 192.016 photometric observations of 165 members of the Themis family of asteroids, finding that 25 objects might exhibit low level cometary activity (the light curves are shown in the SI).

**(b)** Three objects may be nearby, contact binary or very elongated objects (Figure, 10), 1576 Fabiola, 1633 Chimay and 1778 Alfven, however since their study is not the scientific objective of this work, we leave their analysis and confirmation for another occasion.

**(c)** Using theoretical equations of the magnitude of the rotational light curve as a function of observed aspect angle, we derive the shape that a rotational light curve should have vs its location in the orbit for an ellipsoid of semi-axis a:b:c and a>b>c. Using this signature, we identify 31 objects of 165 studied (18.2%), showing evidence of large obliquity (Table 2, Figure 9 and SI).

**(d)** 106 objects do not show evidence of activity (Table 2, Figure 6 and SI). From the dispersion of the data envelope, we deduce a detection limit of ~<±0.20 magnitudes.

**(e)** We show that there is a large systematic error in the absolute magnitude assigned by the Minor Planet Center to asteroids, amounting to a mean value of -0.32±0.19 magnitudes but reaching to ~1.0 magnitudes in some cases. This affects the calculated diameter of these objects (Figure 11).

**(f)** The LAG parameter measures the time with respect to perihelion at which the maximum activity takes place, and a histogram of these values of our candidates for the Themis population (Figure 13), shows that there is a peak just after perihelion, indication of water ice sublimation. It also shows that there are 20 objects, with LAG after perihelion vs 15 before perihelion. This might be interpreted as a thermal lag: the thermal wave has to penetrate below the surface to sublime underlying layers of ice.

**(g)** We find that the duration of activity of the active asteroids of the Themis family is surprisingly long, with many objects active for more than a year (Figure 14).

**(h)** We used the LCDB of Warner et al. (2009) because it provides rotational periods of more than 5000 asteroids. Thus we can test the hypothesis that the activity of some of these objects may be due to rotational instability (Jewitt et al., 2014). We find evidence of an excess of fast rotators (Figure 15), however we do not find evidence that this mechanism is working (Figures 16 and 17).

**(j)** In the phase space Asec vs Diameter, the active members of the Themis family plot in the lower side, with Asec < 1 magnitude. This is the graveyard. Since some members are larger than 100 km in diameter, they also plot in the goliath region (Figure 18).

**(k)** We found an excess of fast rotators. Masiero et al. (2009) also did. We want to advance the hypothesis that fast rotator might the remnant of old comets that have suffered spin up due to jets, but that did not split because there is a monolith at the center.

**(l)** As of September 2016, there are 717768 asteroids listed in the Minor Planet files. If we assume that we do not have any false positives and the percentage of positive objects found in the Themis family can be extrapolated to the whole Main Belt, the number of potentially active asteroid gets to the very large number of ~111.000.

This number is much larger than the ones predicted in previous surveys and indicates one of three scenarios: A) There are many false positives in our detections and the real number of active asteroid is much smaller than we found. This would be tantamount to saying that the MPC astrometric-photometric database is only astrometric and not photometric, invalidating absolute magnitude determinations based on this database. B) The location of active asteroids is restricted to the Themis family and an extrapolation to the whole belt is not possible. Or C) there are few false positives in our candidates and the main belt actually contains many low level active asteroids undetected by current surveys. Case C) would imply that the main belt is not a field of bare rocks but a graveyard of extinct comets, changing our current paradigm of the main belt. So it is of the outmost importance to verify *observationally* our candidates, and determine which of these scenarios is valid.

## 15. Acknowledgements

To an anonymous referee for his contribution to improve the scientific quality of this investigation. The FACom group is supported by the project "Strategy of Sustainability 2015 - 2016", sponsored by the Vicerectoría de Investigación of the University of Antioquia. In order to speed up the processing and verification of so many objects, we enrolled the help of undergraduate students of Astronomy at the University of Antioquia, and assigned this research as a Laboratory exercise. The following 13 students, in order of number of objects processed (after the name), participated in this undergraduate experiment: Andrea Arcila Zuluaga (37); Johnny Agudelo (32); Jefferson Arboleda (21); Sebastián Cano (18); Mario Saldarriaga (14); Paula Leguizamón (10); Julio Sánchez Zapata (10); Harold Patiño (5); Mauricio Medina (5); Julieta Osorio (4); Lina Giraldo (4); Diana Rodríguez (1); Valentín Ospina (1).

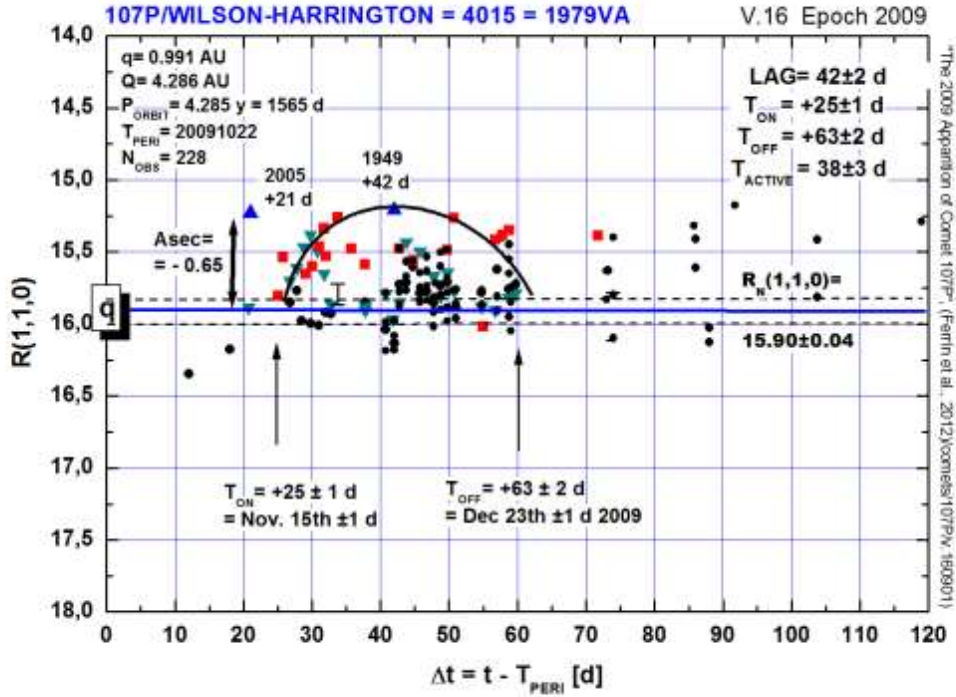

**Figure 1. Comparison light curve.** SLC of comet 107P/Wilson-Harrington. It was discovery on 1979 as an asteroid and received the designation 1979VA = 4015. The vertical axis plots de absolute magnitude R(1,1,0) in the red band. The horizontal axis plots the time with respect to perihelion. The horizontal line is the absolute magnitude $m_R(1,1,0)=15.90\pm0.05$ determined using the phase plot (not shown). Activity started at $T_{ON}$ = ~25±1 d after perihelion, confirmed by two independent data sets (cometas_obs and Ferrín et al. 2012), and reached a maximum at LAG ~+42±2 d, suggesting that the thermal wave took all that time to reach an underneath layer of ice before sublimation started. In the 2009 apparition it never exhibited a coma. The activity enhancement was very shallow (~ -0.65±0.10 magnitudes) and its duration very short (~35±2 days). The candidate objects presented in this work, have very similar light curves. Notice that the sole database of the MPC was able to detect an enhancement for this object.

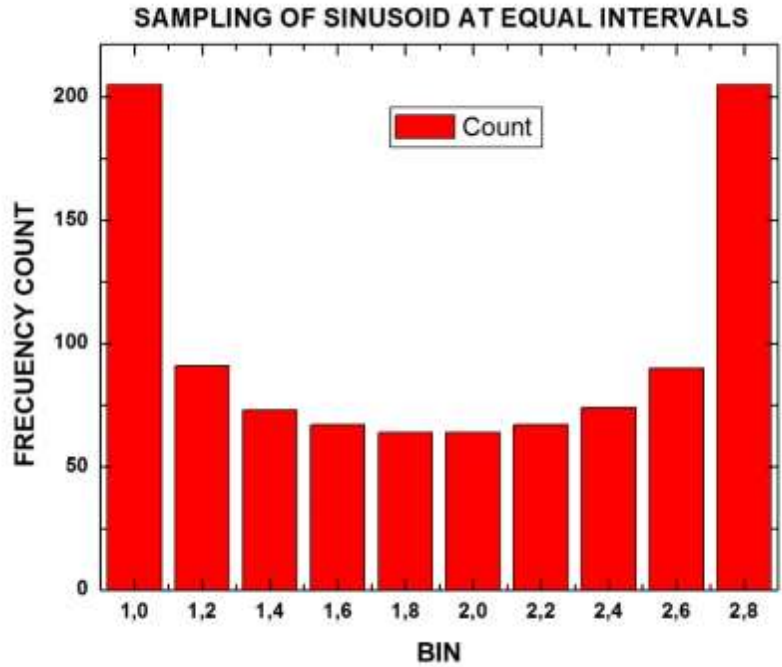

**Figure 2. The rotational light curve of an asteroid** can be modeled as a sine wave. If a sine wave is sampled at identical intervals, the distribution shows maxima at the extremes, the maximum magnitude and the minimum magnitude. This makes the envelope sharper. In order to make the envelope even sharper, we reduced the observations made with each filter to V magnitudes using the Jordi et al. (2008) transformation equations. In this way, our error of detection is around ~<±0.20 magnitudes (confirmation Figure 3).

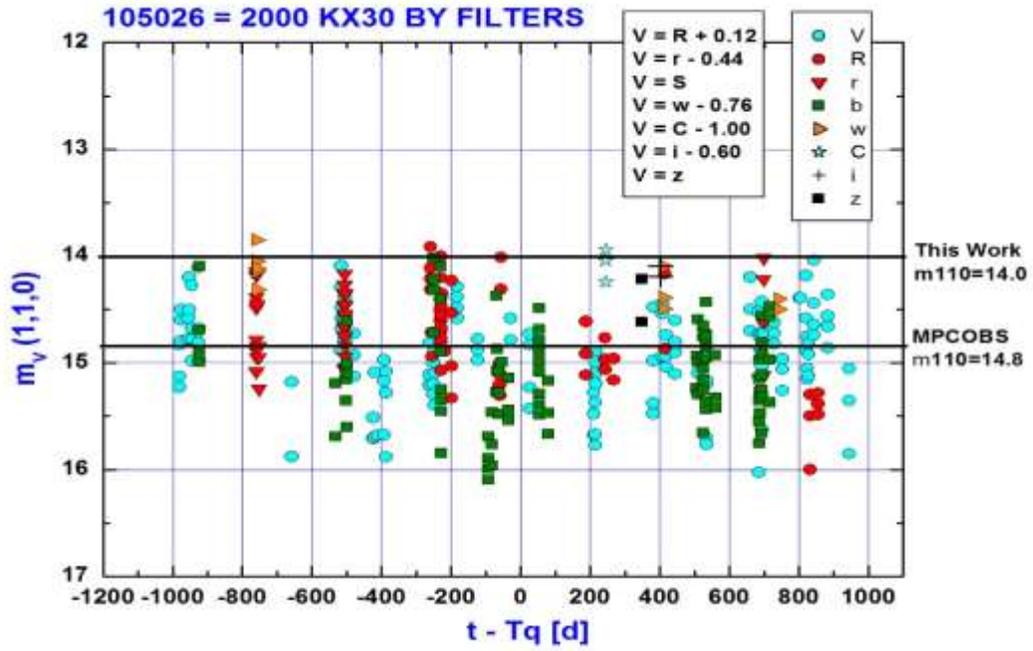

**Figure 3.** The quality of the fit after filter correction can be appreciated in this Figure. The amount of the corrections is listed in the inset frame. All the observations have been reduced to the V-band. Notice the large difference of the absolute magnitude obtained in this work, compared with the value given by the MPC. The scatter of this data is less than $\sim<\pm0.2$ mag.

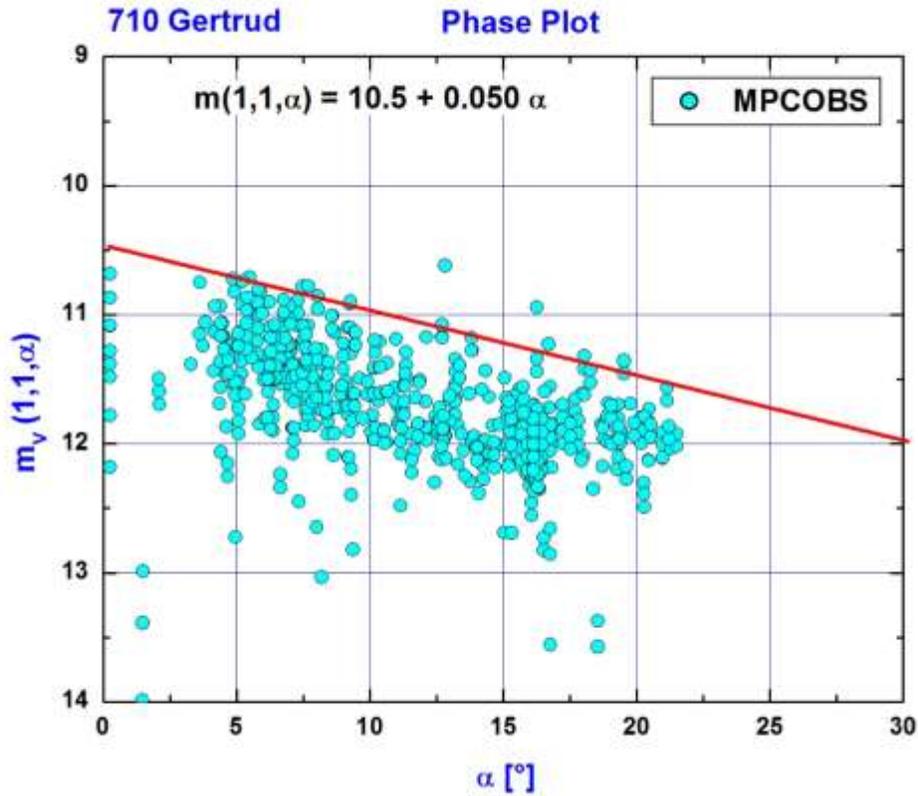

**Figure 4.** The best way to determine the absolute magnitude, is using the phase plot and extrapolating to zero phase angle. The <mean> value of the phase coefficient (the slope), has been measured as $<\beta> = 0.050 \pm 0.002$ for 105 inactive asteroids. The opposition effect ($<5°$) has been removed from these data sets, and thus the enhancements and bumps found in our detections cannot be due to that effect. All objects listed in this work belonging to the Themis family have phase angles $<23°$, thus a linear fit is sufficient to describe the trend.

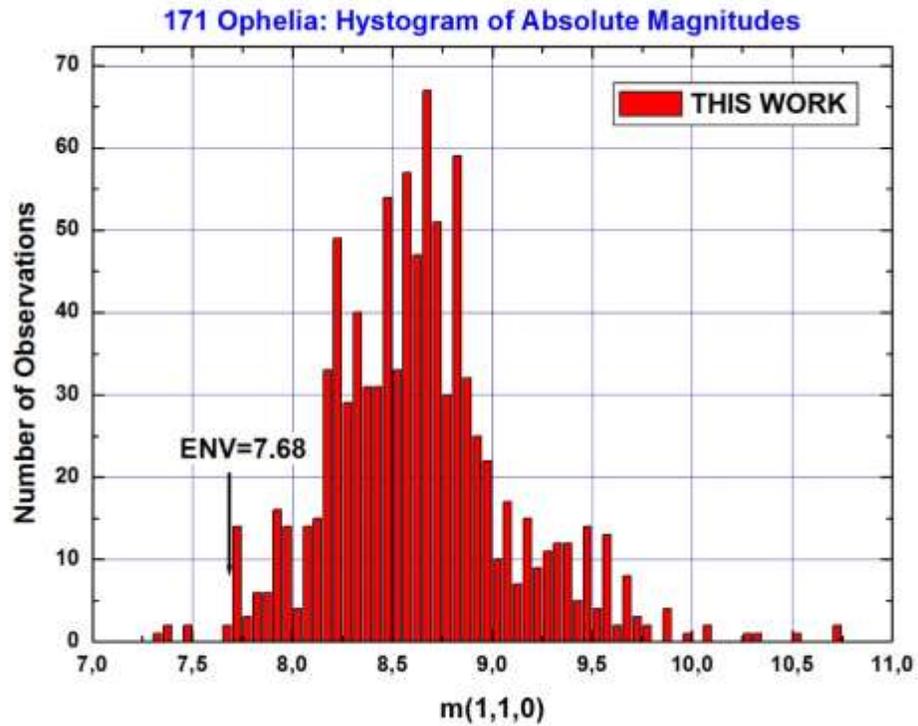

**Figure 5. Another way to determine the absolute magnitude.** The histogram shows the magnitudes distribution. The absolute magnitude can be defined as the brightest edge of the histogram. The agreement with the previous method (Figure 4) is better than ±0.1 mag. However this method cannot be used with active objects because they distort the diagram at the base, faking the envelope.

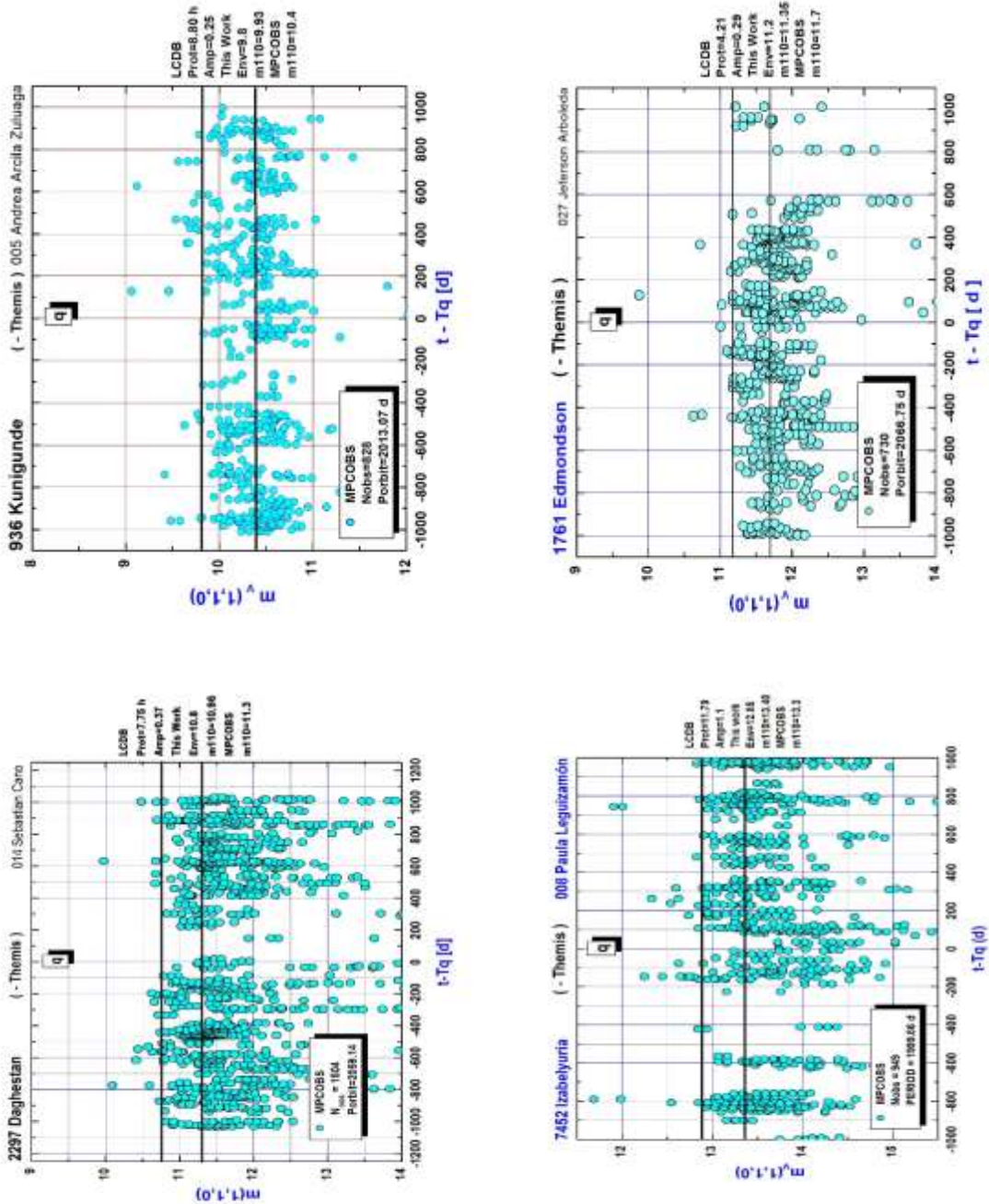

**Figure 6. Four asteroids without activity ( - ).** In the vertical axis, we plot the absolute magnitude $m_V(1,1,0)$ corrected for phase effects. In the horizontal axis, we plot time in the orbit, with respect to perihelion from (–) aphelion to (+) aphelion. The two horizontal black lines show the absolute magnitude found in this work, and that listed in the MPCOBS. Also notice how flat the envelope is, implying that the absolute magnitude of the object is independent of the location in the orbit as required by the definition of absolute magnitude.

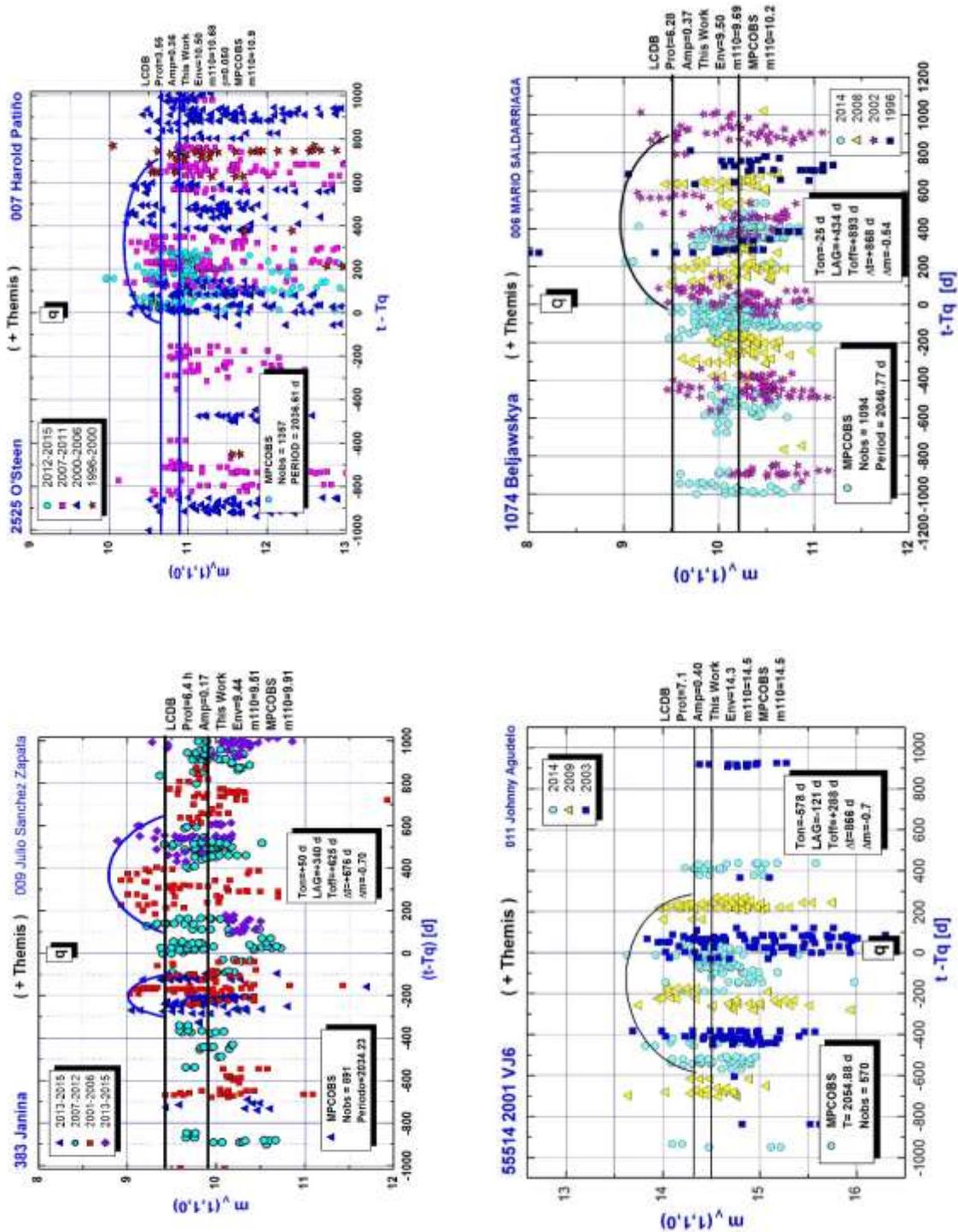

**Figure 7. Four asteroids with activity.** If the activity takes place at the same location in the orbit, at different apparitions, it must be real. All four asteroids fulfil this condition. The other 21 active asteroids belonging to this family are shown in the SI.

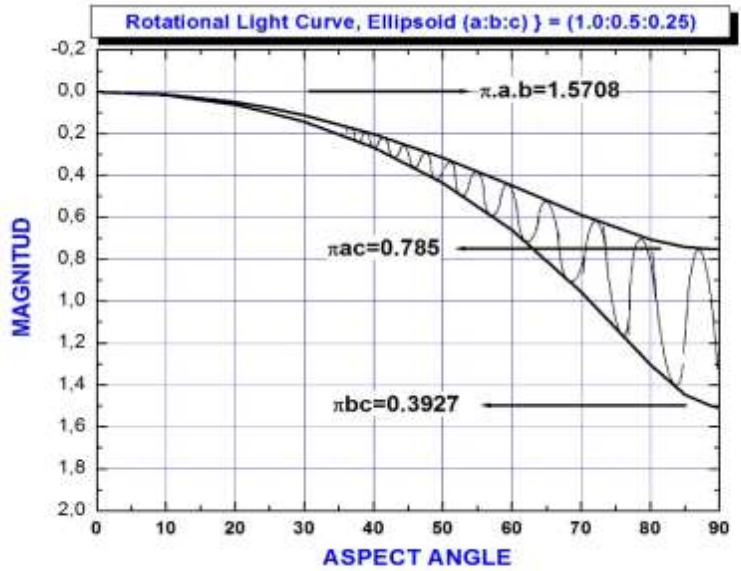

**Figure 8. Theoretical light curve of an ellipsoid** of semi-axis a:b:c vs aspect angle. This is only one fourth of the light curve plotted in a m(1,1,0) vs (–) Aphelion to (+) Aphelion diagram used in this work. When the aspect angle is 0° we see the ellipsoid pole on and the amplitude of the rotational light curve collapses to zero. When the aspect angle is 90° we see the ellipsoid equator on and the amplitude of the rotational light curve is maximum. Thus for large obliquity objects (large aspect angle), the secular light curve will show two maxima and two minima of the envelope, and this signature is easy to detect. However since the subject of this manuscript is not inverting light curves, we leave this analysis undone. Only a few oscillations of the rotational light curve are shown, for clarity. In reality, there are thousands.

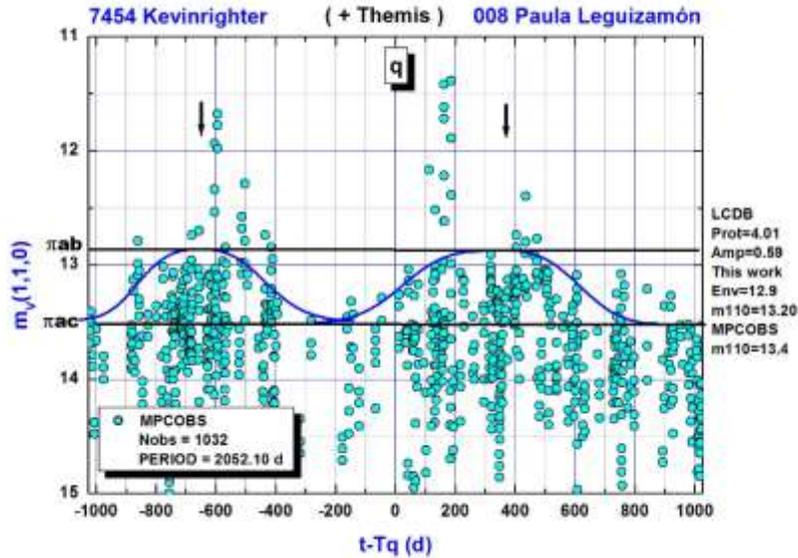

**Figure 9. Asteroid 7454 Kevinrighter** shows evidence of large obliquity. The upper line is the magnitude proportional to log πab, while the lower line is the magnitude proportional to log πac. This signature of large obliquity is relatively easy to detect because it implies two maxima and two minima of the light curve, separated by about half the orbital period (arrows). The other members appear in the SI. These objects deserve further study. The wavy line does not represent a theoretical fit.

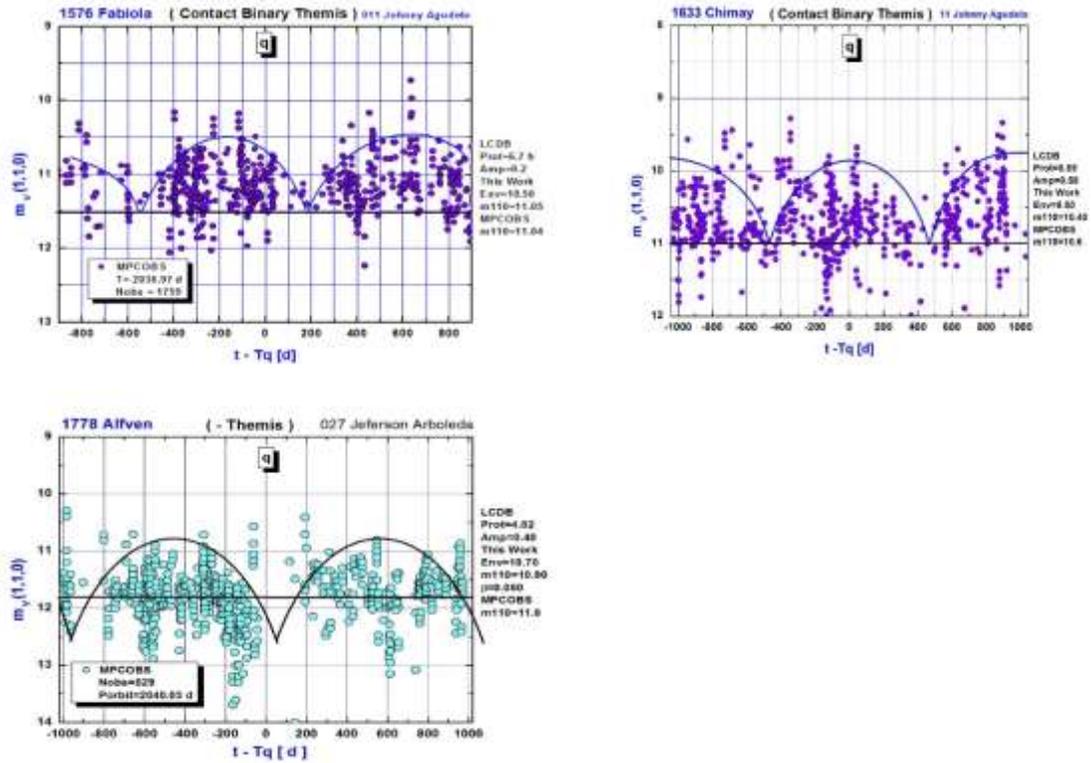

**Figure 10 . Three asteroids 1576 Fabiola, 1633 Chimay, and 1778 Alfven** exhibit the signature of a nearby, contact binary or very elongated object, two maxima and two minima separated by about half the orbital period with the shape of an inverted U (Lacerda, 2011). 1778 Alfven may have the largest amplitude (~1.8 mag) of all previously known contact binaries (confirmation Table 4). These objects should be studied in more detail.

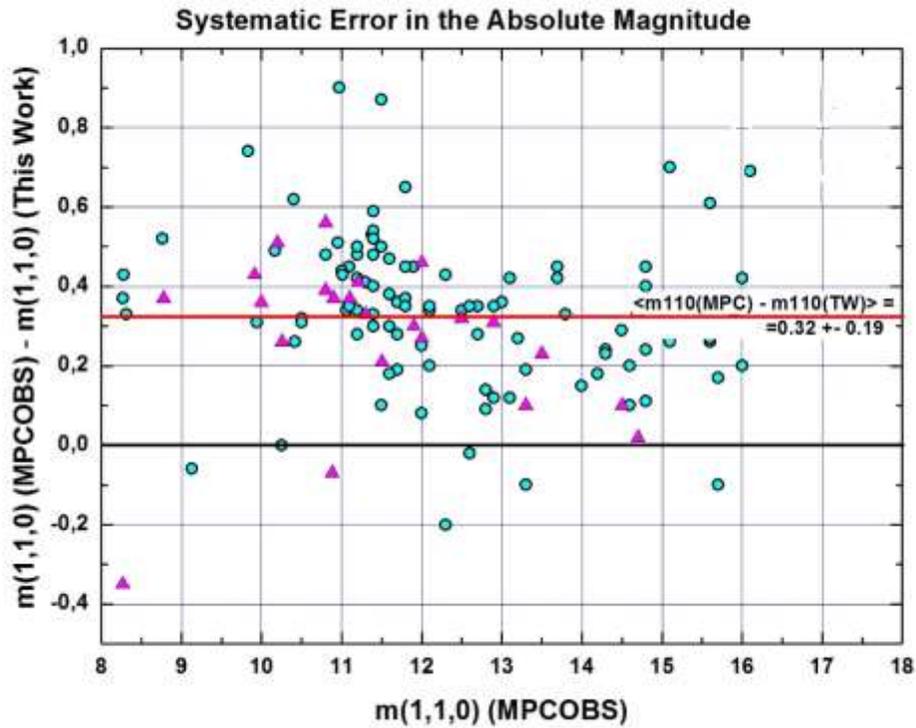

**Figure 11. Systematic difference** between the absolute magnitudes of the MPCOBS and this work. MPCOBS magnitudes are fainter systematically than those measured in this work, by about ~0.32±0.19 magnitudes. Circles are our candidates to inactive asteroids, triangles, our positive candidates. Some deviant object reach almost to 1 magnitude. This is the reason why using MPC absolute magnitudes is not a recommended procedure.

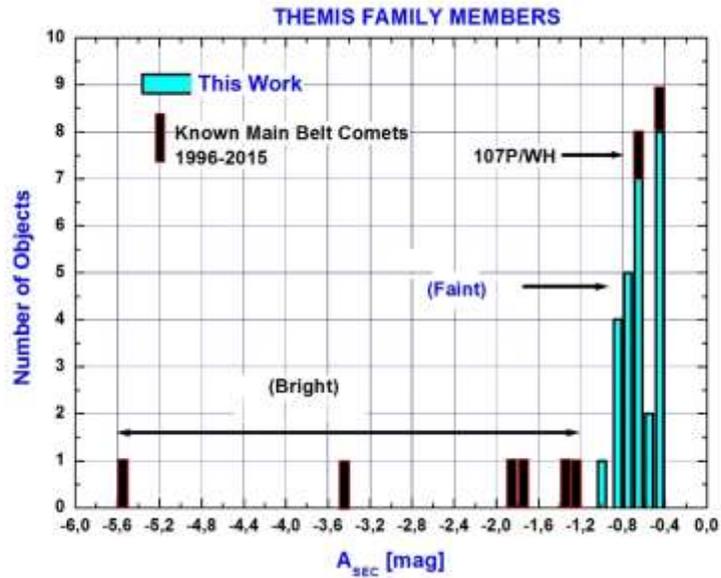

**Figure 12. The main belt comets** discovered up to now (1996-2015) (black vertical rectangles), have a larger amplitude of the SLCs, $A_{SEC}$, than the suspected asteroids of the Themis family, thus they are easy to detect (bright). The objects proposed in this work are difficult to find observationally because their coma is contained inside the seeing disc (faint). The SLCs of these low activity comets resemble that of the well-studied comet 107P/Wilson-Harrington (arrowed, and Figure 1).

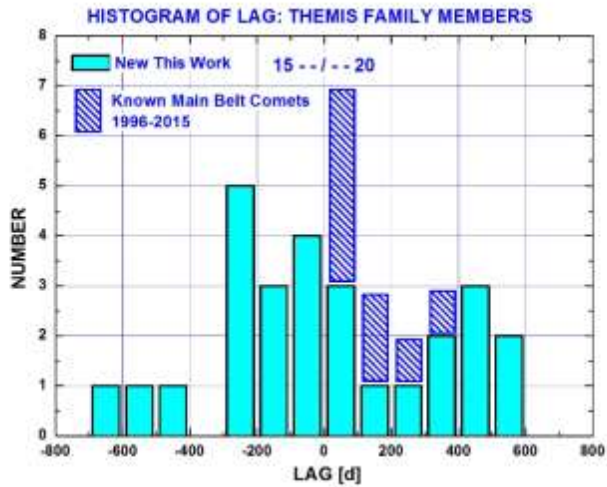

**Figure 13. The main belt comets discovered up to now** (1996-2015) are active near perihelion while the suspected objects in this work seem to be active near and far from perihelion. The fact that 20 objects are located after perihelion vs 15 before perihelion may imply that there is a thermal lag, ice lies deep inside the object, and the thermal wave takes some time to reach that layer. Thus these might not be "ice coated asteroids", but true cometary objects in depth.

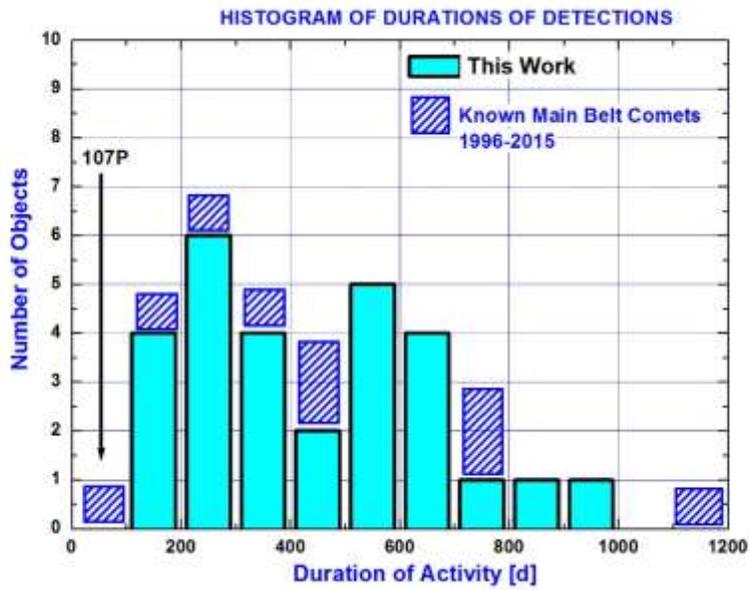

**Figure 14. Duration of activity** of the candidates to active members of the Themis family. Many asteroids show continuous activity during one year or more.

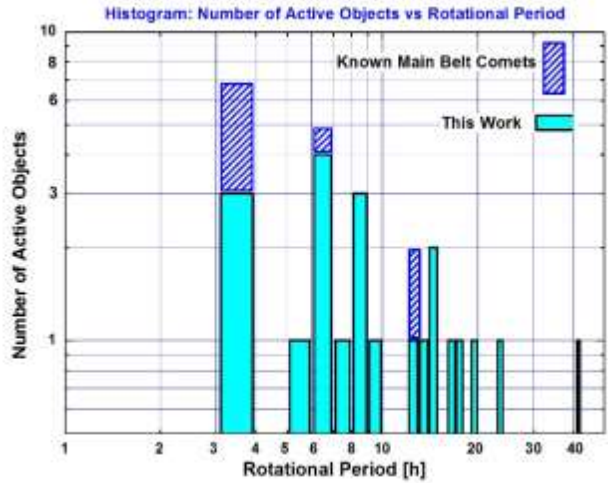

**Figure 15. The histogram shows the number of active comets in the main belt, MBCs, vs their rotational period.** The candidate objects have periods up to 43 hours. The two axis are plotted as logs for clarity. There is an excess of objects with fast rotational periods but their activity is not due to rotational disruption (Jewitt et al., 2015) as deduced from Figures 16 and 17.

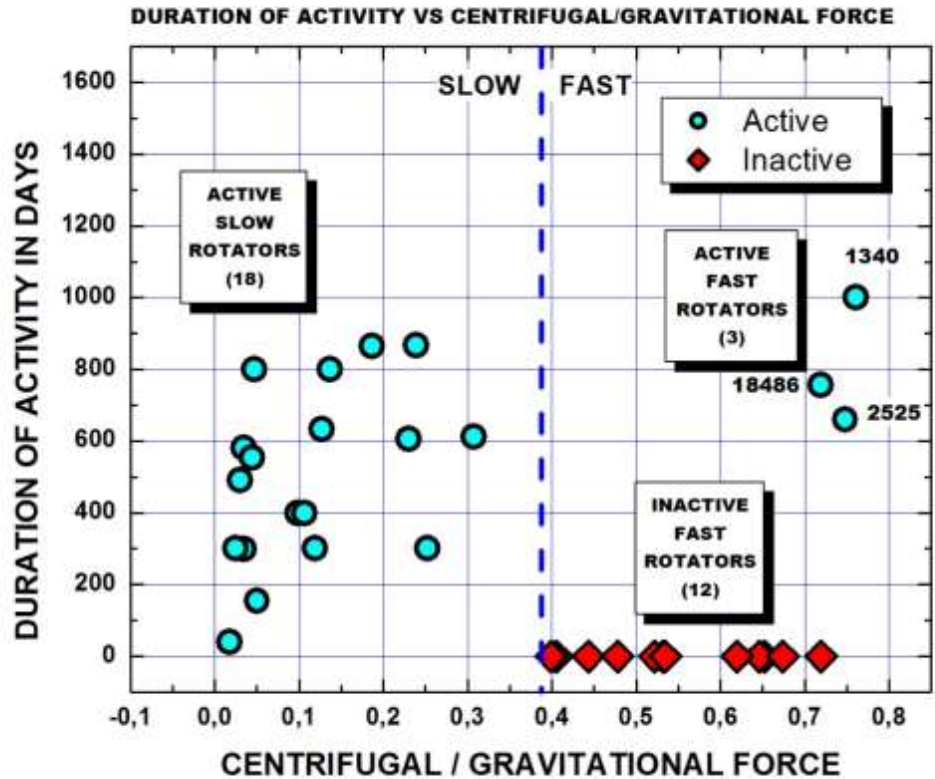

**Figure 16. The duration of activity of the active asteroids is plotted vs the ratio centrifugal/gravitational force** for members of the Themis family. The fact that other fast rotators do not show activity implies that fast rotation is not a sufficient condition for activation. Additionally, active fast rotators have durations similar to those of slow rotators implying that duration does not correlate with rotation.

**Figure 17. The amplitude of the activity in magnitudes for the active objects of the Themis family vs the ratio centrifugal to gravitational force**. The fact that other fast rotators do not show activity implies that fast rotation is not a sufficient condition for activation. Additionally, active fast rotators have amplitudes similar to those of slow rotators implying that amplitude does not correlate with rotation.

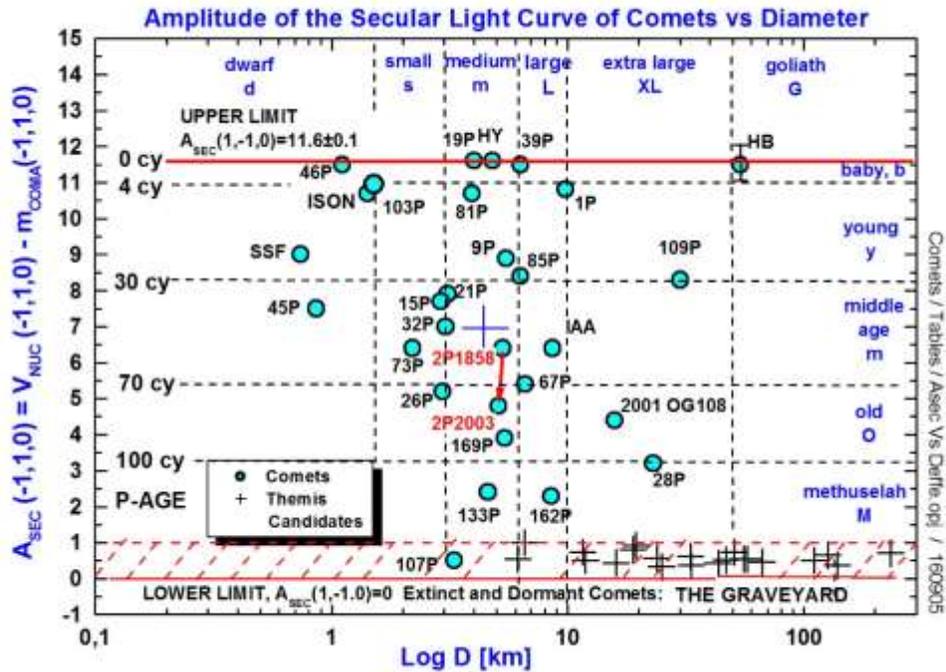

**Figure 18. The location of the candidates to active asteroids of the Themis family in the $A_{SEC}$ vs Diameter diagram.** $A_{SEC}$ measures the activity of the asteroid with respect to the nucleus magnitude. $A_{SEC} = 0$ corresponds to a bare nucleus with no activity. All objects have amplitudes of activity less than one magnitude, and thus lie in the graveyard region of the diagram. Some asteroids are large and also lie in the region of goliath comets. At the top, the envelope suggests that there may be an upper limit to cometary activity.

Table 1.  Orbital elements of selected Themis' family members

| Asteroid or comet | a | e | i | Comments |
|---|---|---|---|---|
| 24 Themis | 3.13 | 0.13 | 0.75° | water ice on surface Campins et al. (2010) Rivkin and Emery (2010) |
| 90 Antiope | 3.15 | 0.17 | 2.20° | water ice on surface Hargrove et al., (2015) |
| 461 Saskia | 3.12 | 0.14 | 1.46° | Hydrated silicates Fornasier et al. (2016) |
| 846 Lipperta | 3.13 | 0.18 | 0.26° | Hydrated silicates Fornasier et al. (2016) |
| 16 Psyche | 2.92 b[2] | 0.13 | 3.1° b[2] | Water or hydroxyl on surface Takir et al. (2017) |
| 133P/Elst-Pizarro | 3.16 | 0.16 | 1.39° | comet |
| 176P/LINEAR | 3.20 | 0.19 | 0.23° | comet |
| 238P/Read | 3.17 | 0.25 b[2] | 1.26° | comet |
| 288P/300163 | 3.05 | 0.20 b[2] | 3.24° b[2] | comet |
| P/2013 R3 Catalina Panstarrs (splitted) | 3.03 b[2] | 0.27 b[2] | 0.86° | comet |

1- According to Zappala et al., (1990), the Themis family is defined by
   $3.05 \leq a \leq 3.22$ AU; $0.12 \leq e \leq 0.19$; $0.7° \leq i \leq 2.22°$.
2- b means that its location is near the border of the family.

Table 2.  Candidates to active, inactive, large obliquity and nearby or contact binary asteroids of the Themis family

| Active Candidates | (2) | (3) | (4) | (5) | (6) | (7) | (8) | (9) | (10) |
|---|---|---|---|---|---|---|---|---|---|
| 24 | 90 | 316 | 379 | 383 | 555 | 561 | 846 | 1027 | 1074 |
| 1253 | 1261 | 1340 | 1438 | 1581 | 1805 | 1956 | 2177 | 2525 | 6221 |
| 8189 | 8976 | 18486 | 55514 | 118826 | | | | | |
| | | | | | | | | | ∑=25 |
| **Inactive Candidates** | | | | | | | | | |
| 62 | 104 | 171 | 222 | 268 | 468 | 526 | 710 | 936 | 946 |
| 954 | 981 | 988 | 996 | 1462 | 1487 | 1489 | 1542 | 1615 | 1623 |
| 1669 | 1674 | 1687 | 1691 | 1761 | 1764 | 1778 | 1782 | 1939 | 2009 |
| 2058 | 2114 | 2165 | 2182 | 2197 | 2203 | 2217 | 2222 | 2228 | 2248 |
| 2250 | 2264 | 2270 | 2296 | 2297 | 2450 | 2464 | 2524 | 2563 | 2627 |
| 2659 | 2708 | 2757 | 2769 | 2781 | 2981 | 3174 | 3186 | 3615 | 3785 |
| 3790 | 3899 | 4242 | 4334 | 4385 | 4741 | 5088 | 5854 | 6353 | 7270 |
| 7452 | 8906 | 10121 | 11776 | 11946 | 16115 | 16228 | 18852 | 27921 | 28895 |
| 30670 | 33764 | 34426 | 38533 | 39339 | 40003 | 57560 | 58541 | 62854 | 65191 |
| 74421 | 105026 | 105047 | 105774 | 113117 | 141240 | 171677 | 173140 | 173974 | 173991 |
| 197402 | 228255 | 241050 | 241908 | 345558 | 197402 | | | | |
| | | | | | | | | | ∑=106 |
| **Candidates To Large Obliquity** | | | | | | | | | |
| 223 | 431 | 492 | 621 | 656 | 938 | 1003 | 1061 | 1073 | 1082 |
| 1259 | 1302 | 1788 | 1815 | 2046 | 2293 | 2324 | 2882 | 2986 | 3525 |
| 3597 | 3618 | 4174 | 4758 | 5036 | 5770 | 7454 | 13533 | 28895 | 32282 |
| 77829 | | | | | | | | | |
| | | | | | | | | | ∑=31 |
| **Candidates to nearby or Contact Binary** | | | | | | | | | |
| 1576 | 1633 | 1778 | | | | | | | ∑=3 |
| | | | | | | | | | ∑=165 |

Table 3. Comets and asteroids with short rotational periods

| **Known objects** | Prot [h] | LAG[1][d] |
|---|---|---|
| 3200 Phaeton | 3.603 | 0 |
| P/2012 F5 Gibbs | 3.24 | +760 |
| (62412) 2000 SY178 | 3.33 | +31 |
| 133P/Elst-Pizarro | 3.471 | +200 |
| **New Objects (This Work)** | | |
| 1340 | 3.52 | -400 |
| 2525 | 3.55 | +290 |
| 18486 | 3.62 | +34 |

1- LAG measures the date of maximum activity with respect to perihelion.
   These values are measured on the SLCs of this work.
   The rotational periods are those of the LCDB (Warner et al., 2009).

Table 4. Candidates to nearby, contact binaries or very elongated objects.

| Object | Group | Diameter [km] | Rotational Period [h] | Δm | Obliquity | Type | References |
|---|---|---|---|---|---|---|---|
| 90 Antiope | Themis Family | 120 | 16.5 | 0.90 | 53° | NB | 1,2 |
| 216 Kleopatra | MBA | 135 | 5.4 | 1.20 | 84° | Bi | 3,4,5 |
| 624 Hektor | Trojan | 230 | 6.9 | 1.10 | 98° | NB | 3,4,6,7 |
| 139775= 2001 QG298 | KBO | 250 | 13.8 | 1.14 | 90° | NB | 7,8,9 |
| 1576 Fabiola | Themis Family | 27.25 | 6.7 | ~1.00 | ~90° | NB, CoB or EO | This Work 10 |
| 1633 Chimay | Themis Family | 36.07 | 6.59 | ~1.16 | ~90° | C NB, CoB or EO | This Work 10 |
| 1778 Alven | Themis Family | 20.51 | 4.82 | ~1.85 | ~90° | ClB, CoB or EO | This Work 10 |

NB = Nearby Binary, CoB=Contact Binary, EO=Elongated Object, Bi=Biloved
References. (1) Michalowsky et al. (2001). (2) Descamps et al. (2009). (3) de Angelis (1995). (4) Kryszczynska et al. (2007). (5) Descamps et al. (2011). (6) Dunlap & Geherls (1969). (7) Lacerda & Jewitt (2007). (8) Sheppard & Jewitt (2004). (9) Lacerda (2011). (10) Warner et al. (2009).